\documentclass[final,numberedheadings]{aipproc}
\layoutstyle{6x9}

\begin{document}

\title{Relativistic Equations for Spin Particles: What Can We Learn From Noncommutativity?}


\author{Valeriy V. Dvoeglazov}{address = {Universidad de Zacatecas, 
Apartado Postal 636, Suc. 3 Cruces,
Zacatecas 98064, Zac., M\'exico }}

\begin{abstract}
We  derive relativistic equations for charged and neutral  spin particles.
The approach for higher-spin particles is based on generalizations of the Bargmann-Wigner  formalism.
Next, we study, what new physical information can the introduction of non-commutativity give us. Additional non-commutative parameters can provide a suitable basis for explanation of the origin of mass.
\end{abstract}

\maketitle

In the spin-1/2 case the Klein-Gordon equation  can be written 
for the two-component spinor ($c=\hbar =1$)
\begin{equation}
(E I^{(2)} - {\mathbf \sigma}\cdot {\bf p})
(E I^{(2)} + {\mathbf \sigma}\cdot {\bf p})\Psi^{(2)} = m^2 \Psi^{(2)}\,,
\end{equation}
or, in the 4-component form
\begin{equation}
[i\gamma_\mu \partial_\mu +m_1 +m_2 \gamma^5 ] \Psi^{(4)} = 0\,.
\end{equation}
There exist various generalizations
of the Dirac formalism, see~\cite{Hua,aaca} and references therein. 
In the higher spin cases  we can proceed in a similar way to obtain relativistic equations.
On this basis we are ready to generalize the BW formalism~\cite{bw-hs}. Why is that convenient? 
In Ref.~\cite{dv-hpa}  the mapping has been presented between the Weinberg-Tucker-Hammer (WTH) equation, Ref.~\cite{WTH}, 
and the equations for antisymmetric tensor  fields (AST). 
\begin{equation}
[\gamma_{\alpha\beta} p_\alpha p_\beta +A p_\alpha p_\alpha +Bm^2 ]
\Psi^{(6)} =0\,,
\end{equation}
which would give many relativistic equations for the AST field differing from the Proca theory.

We tried to find relations between the generalized WTH theory
and other spin-1 formalisms.  Therefore, we were forced to modify the Bargmann-Wigner formalism~\cite{dv-ps}, 
which as has been claimed, does not deal with the parity discrete symmetry. 
For instance, we introduced the sign operator $\epsilon_i$ in the Dirac equations which are the input 
for the formalism for the symmetric 2-rank spinor:
\begin{eqnarray}
\left [ i\gamma_\mu \partial_\mu + \epsilon_1 m_1 +\epsilon_2 m_2 \gamma_5
\right ]_{\alpha\beta} \Psi_{\beta\gamma} &=&0\,,\\
\left [ i\gamma_\mu
\partial_\mu + \epsilon_3 m_1 +\epsilon_4 m_2 \gamma_5 \right ]_{\gamma\beta}
\Psi_{\alpha\beta} &=&0\,,
\end{eqnarray}
In general we have 16 possible combinations, but 4 of them give the same
sets of the Proca-like equations. We obtain~\cite{dv-ps}:
\begin{eqnarray} 
&&\partial_\mu A_\lambda - \partial_\lambda A_\mu + 2m_1 A_1 F_{\mu \lambda}
+im_2 A_2 \epsilon_{\alpha\beta\mu\lambda} F_{\alpha\beta} =0\,,\\
&&\partial_\lambda
F_{\mu \lambda} - \frac{m_1}{2} A_1 A_\mu -\frac{m_2}{2} B_2 \tilde
A_\mu=0\,,
\end{eqnarray} 
with 
$A_1 = (\epsilon_1 +\epsilon_3) /2$,
$A_2 = (\epsilon_2 +\epsilon_4 )/ 2$,
$B_1 = (\epsilon_1 -\epsilon_3 )/ 2$,
and
$B_2 = (\epsilon_2 -\epsilon_4 )/ 2$. See the additional constraints in the cited papers.
So, we have the dual tensor and the pseudovector potential
in the Proca-like sets. The pseudovector potential is the same as that
which enters in the Duffin-Kemmer set for the spin 0. 
Moreover, it appears that the properties of the polarization
vectors with respect to parity operation depend on the choice of the spin basis.
For instance, in Ref.~\cite{dv-ps,GR} the momentum-space polarization vectors have been 
listed in the helicity basis.
Berestetski\u{\i}, Lifshitz and Pitaevski\u{\i} claimed too, Ref.~\cite{BLP}, that
the helicity states cannot be the parity states. If one applies common-used
relations between fields and potentials it appears that the ${\bf E}$ and ${\bf B}$ fields have no usual properties with respect to the space inversion.

Next, we developed the theory of the 4-vector field in the matrix form, including the spin-0 state~\cite{Weinb,dvo-ijmp}.
S. I. Kruglov proposed, Ref.~\cite{krug1}, 
a general form of the Lagrangian for  4-potential field $B_\mu$. 
We have 
\begin{equation}
\alpha \partial_\mu \partial_\nu B_\nu +\beta \partial_\nu^2 B_\mu +\gamma m^2 B_\mu =0\, ,\label{eq-pot}
\end{equation}
provided that derivatives commute.
When $\partial_\nu B_\nu =0$ (the Lorentz gauge) we obtain spin-1 states only.
However, if it is not equal to zero we have a scalar field and an pseudovector potential. 
The consistent theory is, in fact, a generalization of the Stueckelberg formalism~\cite{hpa-stu}.

The spin-2 case has also been considered in a similar way~\cite{dvo-aaca}.
We begin with the equations for the 4-rank symmetric spinor:
\begin{eqnarray}
&&\left [ i\gamma^\mu \partial_\mu - m \right ]_{\alpha\alpha^\prime}
\Psi_{\alpha^\prime \beta\gamma\delta} = 0\, ,
\left [ i\gamma^\mu \partial_\mu - m \right ]_{\beta\beta^\prime}
\Psi_{\alpha\beta^\prime \gamma\delta} = 0\\
&&\left [ i\gamma^\mu \partial_\mu - m \right ]_{\gamma\gamma^\prime}
\Psi_{\alpha\beta\gamma^\prime \delta} = 0\, ,
\left [ i\gamma^\mu \partial_\mu - m \right ]_{\delta\delta^\prime}
\Psi_{\alpha\beta\gamma\delta^\prime} = 0.
\end{eqnarray} 
The massless limit (if one needs) should be taken in the end of all
calculations.

We proceed expanding the field function in the set of symmetric matrices
(as in the spin-1 case). The total function is
\begin{eqnarray}
\lefteqn{\Psi_{\{\alpha\beta\}\{\gamma\delta\}}
= (\gamma_\mu R)_{\alpha\beta} (\gamma^\kappa R)_{\gamma\delta}
G_\kappa^{\quad \mu} + (\gamma_\mu R)_{\alpha\beta} (\sigma^{\kappa\tau}
R)_{\gamma\delta} F_{\kappa\tau}^{\quad \mu} + } \nonumber\\
&+& (\sigma_{\mu\nu} R)_{\alpha\beta} (\gamma^\kappa R)_{\gamma\delta}
T_\kappa^{\quad \mu\nu} + (\sigma_{\mu\nu} R)_{\alpha\beta}
(\sigma^{\kappa\tau} R)_{\gamma\delta} R_{\kappa\tau}^{\quad\mu\nu};
\end{eqnarray}
and the resulting tensor equations are:
\begin{eqnarray}
&&\frac{2}{m} \partial_\mu T_\kappa^{\quad \mu\nu} =
-G_{\kappa}^{\quad\nu}\, ,
\frac{2}{m} \partial_\mu R_{\kappa\tau}^{\quad \mu\nu} =
-F_{\kappa\tau}^{\quad\nu}\, ,\\
&& T_{\kappa}^{\quad \mu\nu} = \frac{1}{2m} \left [
\partial^\mu G_{\kappa}^{\quad\nu}
- \partial^\nu G_{\kappa}^{\quad \mu} \right ] \, ,\\
&& R_{\kappa\tau}^{\quad \mu\nu} = \frac{1}{2m} \left [
\partial^\mu F_{\kappa\tau}^{\quad\nu}
- \partial^\nu F_{\kappa\tau}^{\quad \mu} \right ] \, .
\end{eqnarray}
The constraints are re-written to
\begin{eqnarray}
&&\frac{1}{m} \partial_\mu G_\kappa^{\quad\mu} = 0\, ,\quad
\frac{1}{m} \partial_\mu F_{\kappa\tau}^{\quad\mu} =0\, ,\\
&& \frac{1}{m} \epsilon_{\alpha\beta\nu\mu} \partial^\alpha
T_\kappa^{\quad\beta\nu} = 0\, ,\quad
\frac{1}{m} \epsilon_{\alpha\beta\nu\mu} \partial^\alpha
R_{\kappa\tau}^{\quad\beta\nu} = 0\, .
\end{eqnarray}
However, we need to make symmetrization over these two sets
of indices $\{ \alpha\beta \}$ and $\{\gamma\delta \}$. The total
symmetry can be ensured if one contracts the function $\Psi_{\{\alpha\beta
\} \{\gamma \delta \}}$ with {\it antisymmetric} matrices
$R^{-1}_{\beta\gamma}$, $(R^{-1} \gamma^5 )_{\beta\gamma}$ and
$(R^{-1} \gamma^5 \gamma^\lambda )_{\beta\gamma}$ and equate
all these contractions to zero (similar to the $j=3/2$ case
considered in Ref.~[3b,p.44].  
We  encountered with
the known difficulty of the theory for spin-2 particles in
the Minkowski space.
We explicitly showed that all field functions become to be equal to zero.
Such a situation cannot be considered as a satisfactory one (because it
does not give us any physical information) and can be corrected in several
ways. We  modified the formalism~\cite{dv-ps}. The field function is now presented as
\begin{equation}
\Psi_{\{\alpha\beta\}\gamma\delta} =
\alpha_1 (\gamma_\mu R)_{\alpha\beta} \Psi^\mu_{\gamma\delta} +
\alpha_2 (\sigma_{\mu\nu} R)_{\alpha\beta} \Psi^{\mu\nu}_{\gamma\delta}
+\alpha_3 (\gamma^5 \sigma_{\mu\nu} R)_{\alpha\beta}
\widetilde \Psi^{\mu\nu}_{\gamma\delta}\, .
\end{equation}
The equations and constraints have been found between tensors of different parity 
properties\cite{dvo-aaca}.

The questions of "non-commutativity" see, for instance, in Ref.~\cite{Bled}.
The assumption that operators of
coordinates do {\it not} commute $[\hat{x}_{\mu },\hat{x}_{\nu }]_{-} = i\theta_{\mu\nu}$ (or, alternatively, $[\hat{x}_{\mu },\hat{x}_{\nu }]_{-}= iC_{\mu\nu}^\beta x_\beta$)
has been first made by H. Snyder~\cite{snyder}. Later it was shown that such an anzatz may lead to non-locality. Thus, the Lorentz symmetry may be
broken. On the other hand, the famous Feynman-Dyson proof of Maxwell equations~\cite{FD}
contains intrinsically the non-commutativity of velocities. While $[ x^i, x^j ]_-=0$ therein, but  $[ \dot x^i (t), 
\dot x^j (t) ]_- = \frac{i\hbar}{m^2} \epsilon^{ijk} B_k \neq 0$ (at the same time with $[x^i, \dot x^j]_- = \frac{i\hbar}{m} \delta^{ij}$) that also may be considered as a contradiction with
the well-accepted theories. Dyson wrote in a very clever way about this problem.
Furthermore, it has recently been shown that notation and terminology, 
which physicists used when speaking about partial
derivative of many-variables functions, are sometimes 
confusing~\cite{eld}. The well-known physical example of the situation, when we have both explicite and implicite dependences of the function which derivatives act upon, is the field of an accelerated charge~ \cite{landau}.
First, Landau and Lifshitz wrote that the functions depended on the retarded time $t^{\prime }$
and only through $t^{\prime }+R(t^{\prime })/c=t$ they depended implicitly
on $x,y,z,t$. However, later they used
the explicit dependence of $R$ and fields on the space coordinates of the
observation point too. Otherwise, the ``simply" retarded fields do not satisfy the Maxwell 
equations. So, actually
the fields and the potentials are the functions of the following forms:
$A^\mu (x, y, z, t' (x,y,z,t)), {\bf E} (x, y, z, t' (x,y,z,t)), {\bf B} (x, y, z, t' (x,y,z,t))$. 

Let us to work out one example in the momentum representation. In the general case 
of the ``whole-partial" derivative one has 
\begin{equation}
{\frac{\hat\partial f ({\bf p}, E({\bf p})) }{\hat\partial p_i}} \equiv {
\frac{\partial f ({\bf p}, E({\bf p})) }{\partial p_i}} + {\frac{\partial f (
{\bf p}, E({\bf p})) }{\partial E}} {\frac{\partial E}{\partial p_i}}\, .
\end{equation}
Applying this rule, we surprisingly find 
\begin{eqnarray}
&&[{\frac{\hat\partial }{\hat\partial p_i}},{\frac{\hat\partial }{\hat
\partial E}}]_- f ({\bf p},E ({\bf p})) = - {\frac{\partial f }{\partial E}} {\frac{\partial
}{\partial E}}({\frac{\partial E}{\partial p_i}})\,.  \label{com}
\end{eqnarray}
We put forward the
following anzatz in the momentum representation: 
\begin{equation}
[\hat x^\mu, \hat x^\nu]_- = \omega ({\bf p}, E({\bf p})) \,
F^{\mu\nu}_{\vert\vert}{\frac{\partial }{\partial E}}\,.
\end{equation}
In the modern literature, the idea of the broken Lorentz invariance by this method is widely discussed. 
Let us turn now to the application of the presented ideas to the Dirac case.
Recently, we analized Sakurai-van der Waerden method of derivations of the Dirac
(and higher-spins too) equation. We can start from either the equation (1) or the 4-component 
equation
\begin{equation}
(E I^{(4)}+{\mathbf \alpha}\cdot {\bf p} +m\beta) (E I^{(4)}-{\mathbf\alpha}\cdot
{\bf p} -m\beta ) \Psi_{(4)} =0.\label{f4}
\end{equation}
We also postulate the non-commutativity
\begin{equation}
[E, {\bf p}^i]_- = \Theta^{0i} = \theta^i,,
\end{equation}
as usual. Therefore the equation (\ref{f4}) will {\it not} lead
to the well-known equation $E^2 -{\bf p}^2 = m^2$. Instead, we have
\begin{equation}
\left \{ E^2 - E ({\mathbf \alpha} \cdot {\bf p})
+({\mathbf \alpha} \cdot {\bf p}) E - {\bf p}^2 - m^2 - i {\mathbf\sigma}\times I_{(2)}
[{\bf p}\otimes {\bf p}] \right \} \Psi_{(4)} = 0
\end{equation}
For the sake of simplicity, we may assume the last term to be zero. Thus we come to
\begin{equation}
\left \{ E^2 - {\bf p}^2 - m^2 -  ({\mathbf \alpha}\cdot {\mathbf \theta})
\right \} \Psi_{(4)} = 0\,.
\end{equation} 
However, let us make the unitary transformation. It is known~\cite{Berg}
that one can
\begin{equation}
U_1 ({\mathbf \sigma}\cdot {\bf a}) U_1^{-1} = \sigma_3 \vert {\bf a} \vert\,.\label{s3}
\end{equation}
The final equation is
\begin{equation}
[E^2 -{\bf p}^2 -m^2 -\gamma^5_{chiral} \vert {\bf \theta}\vert ] \Psi^\prime_{(4)} = 0\,.
\end{equation}
In the physical sense this implies the mass splitting for the Dirac particle over the non-commutative space. We have two solutions for $m_1 =\sqrt{m^2 +\vert \theta\vert}$ and $m_2 =\sqrt{m^2 -\vert \theta\vert}$. This procedure may be attractive for explanation of the mass creation and the mass splitting for fermions.

The conclusions are: 1) The $(1/2,1/2)$ representation contains both the  spin-1 and spin-0
states (cf. with the Stueckelberg formalism). \, 2)
Unless we take into account the fourth state (the ``time-like" state, or
the spin-0 state) the set of 4-vectors is {\it not} a complete set in a mathematical sense.\, 
3) We cannot remove terms like $(\partial_\mu B^\ast_\mu)(\partial_\nu B_\nu)$ 
terms from the Lagrangian and dynamical invariants unless apply the Fermi 
method, i.~e., manually. The Lorentz condition applies only to the spin 1 states.\,
4) We have some additional terms in the expressions of the energy-mo\-men\-tum vector (and, accordingly, of the 4-current and the Pauli-Lunbanski vectors), which are the consequence of the impossibility to apply the Lorentz condition for spin-0 states.\,
5) Helicity vectors are not eigenvectors of the parity operator. Meanwhile, the parity is a ``good" quantum number, $[{\cal P}, {\cal H}]_- =0$ in the Fock space.\,
6) We are able to describe the states of different masses in this representation from the beginning.\,
7) Various-type field operators can be constructed in the $(1/2,1/2)$ representation space. For instance, they can contain $C$, $P$ and $CP$ conjugate states.
Even if $b_\lambda^\dagger =a_\lambda^\dagger$ 
we can have complex 4-vector fields.
We found the relations between creation, annihilation operators for different types of the field operators $B_\mu$.\, 8)  Propagators have good behavious in the massless limit as opposed to those of the Proca theory.\, 9)  The spin-2 case can be considered on an equal footing with the spin-1 case.\, 
10)  The postulate of non-commutativity leads to the mass spliting for leptons.

\begin{theacknowledgments}
  It is a pleasure to thank the organizers and participants of the  Workshop on
  Geometric Methods in Physics in Bia{\l}owie\.{z}a. Special thanks to
  Prof. Anatol~Odzijewicz for the  hospitality. 
\end{theacknowledgments}

\end{document}